# Toward a Dynamical Taxonomy of Insomnia:

## A Multiaxial Framework for Sleep-State Transitions and Architectural Failure

Alexander Poltorak

NeuroLight, Inc. | Pomona, NY 10970

Corresponding author: Alex@NeuroLight.co

## Abstract

Insomnia disorder is defined at the syndrome level, yet similar complaints can arise from different abnormalities in sleep regulation, state transition, state stabilization, spatial recruitment, architectural sequencing, and state perception. We propose a multiaxial dynamical framework whose principal contribution is organizational: a candidate profile is specified by the dynamical operation that fails, the sleep stage or boundary at which it fails, and its causal status. Objective sleep duration, age, circadian phase, comorbidity, medication exposure, and night-to-night variability are modifier/covariate dimensions rather than additional mechanistic classes. A local Landau-Ginzburg formalism, adopted from prior cortical and sleep-dynamics work, supplies a phenomenological language for nested hypotheses. Its relaxational form applies only to boundary-local dynamics under an approximate gradient description; non-gradient escape requires an action or quasipotential treatment, and whole-night REM-NREM sequencing requires reactive or oscillatory dynamics. Routine polysomnography usually identifies effective combinations rather than curvature, escape action, bias, noise, and relaxation separately. The strongest boundary-level evidence concerns sleep onset, where published results are consistent with bifurcation-like or bistable dynamics but do not yet exclude a driven smooth transition produced by the homeostatic-circadian ramp. The remaining operation classes are hypothesis-generating extensions. The taxonomy is judged by pragmatic utility, including improved communication, stratification, and prediction; the scalar-field implementation and specific dynamical profiles are separately falsifiable. The framework is a phenomenological organizing model rather than a new diagnosis, validated biomarker, or treatment-selection system.

**Keywords**: insomnia; sleep architecture; dynamical systems; Landau-Ginzburg; Landau phenomenology; two-process model; sleep-stage transitions; phase-transition-like dynamics; hyperarousal; slow-wave sleep; REM instability

## Key points

- Insomnia is organized along three axes: dynamical operation, stage or boundary, and causal status; established phenotypes and severity variables are retained as modifier/covariate dimensions.

- The relaxational Landau-Ginzburg model is a boundary-local approximation. Gradient potential barriers, non-gradient escape actions, local recovery, fluctuation intensity, and spatial coupling must be distinguished and require different data.
- Sleep-onset findings provide the strongest boundary-level evidence, but they are not yet mechanism-identifying because a driven Process-*S*/Process-*C* trajectory can mimic passive early-warning signals; the other operations remain hypothesis-generating extensions.
- The taxonomy earns its place only through pragmatic utility beyond conventional subtyping. The scalar coordinate, Landau-Ginzburg implementation, and individual mechanistic profiles remain separately falsifiable.

# 1. Introduction

Insomnia disorder is common, clinically important, and physiologically heterogeneous. Current diagnostic systems define it by persistent difficulty initiating sleep, maintaining sleep, waking earlier than intended, or obtaining sleep experienced as poor or nonrestorative, together with daytime consequences and adequate opportunity for sleep [1,9]. This syndrome-level definition appropriately prioritizes distress and impairment, but it aggregates patients whose complaints may arise through different physiological routes.

Heterogeneity appears in cortical, autonomic, and metabolic arousal [11-20]; NREM spectral and microstructural organization [21-22,29]; sleep-state perception [23-28]; REM organization [33,34]; and objective sleep duration and persistence [36-39]. Group-level abnormalities are inconsistent across cohorts, which argues against treating hyperarousal, reduced slow-wave sleep, or any single measure as a universal defect.

Existing models explain important parts of this heterogeneity. The two-process model describes slow homeostatic and circadian control [2-4]; flip-flop circuitry explains rapid state switching [5-7]; the 3P model organizes predisposing, precipitating, and perpetuating factors [8]; hyperarousal models describe persistent cognitive, cortical, autonomic, or metabolic activation [10]; and state-space, transition, and field models quantify or generate sleep dynamics [43-49]. These approaches do not provide a common classification of whether a complaint reflects failure to approach a transition, cross it, stabilize the destination state, recruit spatial organization, preserve architectural sequencing, or map physiological sleep onto scored and experienced sleep.

## 1.1 Central hypothesis and scope

The central hypothesis is that insomnia complaints may arise from separable abnormalities in slow control trajectories, state-boundary crossing, post-transition stabilization, spatial recruitment, architectural sequencing, and the mapping between physiological and experienced sleep. A low-dimensional nonequilibrium framework can organize these abnormalities and generate testable distinctions that are not captured by symptom labels or conventional stage totals alone.

This article presents a conceptual framework with a phenomenological dynamical model. Its taxonomy classifies operation, location, and causal status. Curvature, barrier height, bias, noise, and related quantities are nested mechanistic candidates within an operation rather than seven coequal patient labels. The framework is not a mechanistic circuit model, a replacement for ICSD diagnosis, a validated diagnostic biomarker, or a treatment-selection system. Circadian rhythm sleep-wake disorders, sleep-disordered breathing, movement disorders, pain, psychiatric illness, medication or substance effects, and neurodegenerative disease remain differential diagnoses or comorbid causes. The framework is intended as an overlay that asks where and how the sleep process fails after those causes are assessed.

The framework makes two principal affirmative predictions. First, local recovery from small perturbations and escape under larger perturbations will dissociate reproducibly across patients and will predict perturbation vulnerability or outcome beyond a single undifferentiated instability measure. Second, rounded-bifurcation or bistable/escape models will predict subject-held-out sleep-onset trajectories better than a Process-*S*/Process-*C*-driven smooth sigmoid. Failure of either prediction would materially weaken the dynamical implementation, while leaving the descriptive ontology to be judged by its independent utility criteria.

## 1.2 Evidence selection

The literature synthesis is purposive and domain-scoping rather than systematic or confirmatory. Studies were selected to identify empirical analogues, boundary conditions, contradictory findings, and plausible falsifiers for each proposed operation, with priority given to primary human PSG/EEG studies, perturbation studies, multi-night recordings, large cohorts, and foundational physiological or computational models. The evidence should not be interpreted as exhaustive or as a quantitative estimate of effect size. At the time of manuscript preparation, references 30, 31, 33, 40, and 41 were peer-reviewed articles; reference 32 was an accepted journal manuscript; and references 42 and 59 were preprints. Preprints are used only for speculative or methodological extensions and carry no essential evidentiary burden.

## 1.3 Candidate dynamical endotype

In this paper, a candidate dynamical endotype is a reproducible within-person pattern of latent or fitted dynamical quantities, transition statistics, or perturbation responses that (i) persists across at least two nights or independent recording conditions, (ii) is distinguishable from conventional stage totals and symptom labels, and (iii) predicts an external criterion such as perturbation vulnerability, clinical course, relapse, treatment response, or subjective-objective discrepancy. Before these criteria are met, the more cautious term mechanistic phenotype should be used.

## 2. Relationship to existing sleep and insomnia models

Process *S* and Process *C* define a slowly varying trajectory through physiological control space. Their gradual evolution can move the system toward a local loss of stability or a barrier-crossing region, after which an abrupt transition occurs [2-4]. Reciprocal inhibition between sleep-promoting and wake-promoting populations supplies a plausible circuit implementation of bistability and rapid switching [5-7]. These models identify the regulators and circuits; the present framework asks which macroscopic operation fails in a particular patient and at a particular boundary.

*Table 1. Relationship of the proposed framework to established models.*

| Existing framework | What it explains | What it does not distinguish | Added value of the proposed framework |
|---|---|---|---|
| Two-process model | Slow timing, sleep propensity, and homeostatic intensity | Local transition geometry and stability after entry | Separates failure to approach, cross, or stabilize a state |
| Flip-flop switch | Circuit basis of rapid sleep-wake switching | Patient-level heterogeneity, spatial recruitment, and observation mapping | Classifies the macroscopic transition pattern implemented by the circuit |
| Hyperarousal and 3P models | Arousal burden, conditioning, and perpetuation | Curvature versus barrier versus bias versus effective noise | Generates distinct perturbation and recovery predictions |
| Transition matrices / HMMs | Empirical transition probabilities and latent states | Physiological interpretation of altered transitions | Links dwell and switching patterns to control, recovery, and escape |
| Neural-mass / field models | Mechanistic population dynamics and whole-cycle behavior | A clinically organized insomnia classification | Provides a low-dimensional clinical organization of dynamical abnormalities |
| Objective short-sleep phenotype | Severity, cardiometabolic risk, and biological burden | Which transition or state-space operation produces short sleep | Treats objective duration as a modifier/covariate dimension across profiles |

The Landau-Ginzburg machinery is adopted from prior work on cortical collective dynamics, sleep-wake transitions, and dynamic critical phenomena [43,45,61]. The paper's original contribution is the three-axis clinical-taxonomic organization and its insistence that operations, locations, and causal roles be classified on separate dimensions. The framework does not claim a new circuit theory of insomnia. It separates control-path abnormalities from boundary-crossing abnormalities, destination-state stabilization failures, spatial-recruitment failures, architectural-sequencing failures, and observation-layer dissociation within one set of testable concepts. Objective short sleep duration, chronicity, and hyperarousal burden can cut across several such profiles [36-39].

The level-mixing problem addressed here does not arise from the formal ICSD-3 diagnostic structure. It arises in the informal phenotype literature, where short sleep, hyperarousal, sleep-state misperception, circadian bias, and architectural abnormalities are sometimes discussed as parallel types even though they represent severity, mechanism, observation, timing, or architecture. The proposed scheme also differs from RDoC, which organizes transdiagnostic functional constructs across units of analysis, and from the former DSM-IV multiaxial system, which distributed clinical disorders, personality and developmental disorders, medical conditions, psychosocial factors, and global functioning across separate diagnostic registers [65,66]. Here, the axes combine a dynamical operation, its sleep-state location, and the causal status of a candidate physiological abnormality within an insomnia-focused research ontology.

The dynamical level earns its place only when it answers predictive questions that behavioral or syndrome-level descriptions leave unresolved. Examples include whether identical sleep-onset latency reflects distant control-path position, delayed boundary loss, or unstable early NREM; whether a patient resists small perturbations yet escapes readily under larger ones; and whether two treatments that produce the same total sleep time alter different transition mechanisms. The taxonomy is an ontology, so it is judged pragmatically rather than as true or false: it should improve inter-rater communication, trial stratification, or prediction beyond onset/maintenance/terminal-awakening labels. The associated dynamical models must additionally beat 3P, hyperarousal, transition-matrix, hidden-state, model-free perturbation, and simpler trajectory baselines in preregistered held-out comparisons.

# 3. Minimal phenomenological dynamical framework

## 3.1 Scope, normalization, and local quasipotential

The sleeping brain is spatially extended, noisy, dissipative, and far from thermodynamic equilibrium. The term phase-transition-like is used operationally for collective reorganization with signatures such as loss of local stability, metastability, bimodality, matched-control path dependence, critical slowing, spatial correlation growth, or nucleation. A scored stage is not automatically an attractor. The term attractor is reserved for an empirically supported stable region of latent state space toward which trajectories return after perturbation.

Let $\phi(r,t)$ in [0,1] denote a dimensionless latent composite NREM-ordering coordinate at cortical position $r$ and time $t$. Larger values indicate greater slow-oscillation organization, lower signal complexity, and broader spatial coordination; these observables are indicators rather than interchangeable definitions. The centered coordinate is $\psi(r,t) = \phi(r,t) - \phi_0$, where $\phi_0$ is a prespecified reference level. The cortical domain $\Omega$ is treated as a one- or two-dimensional effective sheet for local analysis. Periodic or no-flux spatial boundary conditions may be used for estimation or simulation and must be reported; neither is asserted to be a literal anatomical boundary.

A local gradient approximation uses the effective quasipotential functional

$$F[\psi;\lambda] = \int_{\Omega} d^d r \; \frac{a(\lambda)\psi^2}{2} + \frac{b(\lambda)\psi^4}{4} + \frac{c\psi^6}{6} + \frac{\kappa|\nabla\psi|^2}{2} - h(\lambda)\psi \tag{1}$$

The coefficient $a$ contributes to local curvature, $b$ sets the leading nonlinear transition character, $c>0$ stabilizes large amplitudes when $b<0$, $\kappa$ penalizes spatial inhomogeneity, and $h$ is a generally nonzero bias or tilt. Sleep states are not symmetry-related phases, so the even polynomial is a centered local normal form rather than a biological symmetry claim. For most boundaries, the simpler $c=0$, $b>0$ sector is sufficient. The sixth-order term is retained only to permit high-amplitude stabilization when $b$ becomes negative and to represent a possible mixed or first-order-like low-order/high-order coexistence regime.

The number and type of stable states are determined by $V'(\psi)=0$ and $V''(\psi)>0$. With $b>0$, $c=0$, and $h$ near zero, crossing $a=0$ produces an idealized continuous transition; finite $h$ rounds it into a crossover. With $b>0$, $a<0$, and a sufficiently small field, two local minima can coexist, and a fold occurs when a minimum and intervening saddle coalesce, $V'=V''=0$. With $b<0$ and $c>0$, the polynomial can support coexistence between low- and high-ordering states. These regimes are hypotheses to be selected by data, not labels assigned from stage names.

## 3.2 Dynamics and noise assumptions

The corresponding stochastic time-dependent Ginzburg-Landau equation is

$$\begin{aligned}\frac{\partial\psi}{\partial t} &= -\Gamma\frac{\delta F}{\delta\psi} + \eta(\mathbf{r},t) \\ &= -\Gamma\left[a\psi + b\psi^3 + c\psi^5 - \kappa\nabla^2\psi - h\right] + \eta(\mathbf{r},t)\end{aligned} \tag{2}$$

$\Gamma$ is a relaxation coefficient. The most general effective noise may be spatially correlated, temporally colored, state dependent, or multiplicative. A minimal additive model can be written as

$$\langle\eta(\mathbf{r},t)\eta(\mathbf{r}',t')\rangle = 2D\,C_r(\mathbf{r},\mathbf{r}')\,C_t(t,t') \tag{3}$$

where $D$ is an effective fluctuation scale and $C_r$ and $C_t$ specify spatial and temporal covariance; white noise is the special case $C_r=\delta(r-r')$ and $C_t=\delta(t-t')$. Measurement and scoring errors belong in the observation model, not in $\eta$. Increased symmetric additive noise in a fixed landscape promotes switching in both directions. Selective return to wake additionally requires a wake-biased field, asymmetric or state-dependent noise, directional coupling to sensory or autonomic perturbations, asymmetric boundaries, or a changing control trajectory.

## 3.3 Gradient versus non-gradient scope

Equation 2 is a Model-A relaxational idealization for a local transition window [61]. With symmetric dissipative mobility and additive white noise, the drift is a functional gradient and the stationary distribution has a Gibbs-like form. In that restricted regime, a height difference in $F$ or $V$ can enter a Kramers-type escape approximation. The paper does not assume that this description is valid for the complete sleep cycle or for every stage boundary.

More generally, sleep dynamics may contain rotational probability currents, adaptation, delayed feedback, respiratory or autonomic forcing, and colored, state-dependent, or multiplicative noise. Escape is then governed by a Freidlin-Wentzell quasipotential or minimum action, denoted *W*, rather than by the ordinary height of *V*; the two coincide only under the gradient/additive-noise approximation [62,63]. We therefore use $\Delta V$ for a gradient potential barrier and $\Delta W$ for a general effective escape action. Quantitative barrier language is restricted to boundaries for which residual-current tests and model comparison support approximate relaxational dynamics. Elsewhere, the operational quantities are empirical escape hazard, minimum-action estimates, and perturbation response.

## 3.4 Curvature, escape action, relative preference, bias, and relaxation

Five properties must remain distinct. Local curvature or stiffness $\mu = V''(\psi_{eq})$ controls the response to sufficiently small perturbations. Under the gradient approximation, $\Delta V = V(\text{saddle}) - V(\text{minimum})$ is the local potential barrier. In a non-gradient system, the corresponding escape quantity is the minimum action or quasipotential difference $\Delta W$. Relative well depth is meaningful only within an approximately gradient representation; more generally, state preference is inferred from occupancy under matched controls and directional probability currents. The field *h* tilts a local gradient model, while $\Gamma$ converts local curvature into a recovery timescale. A state can therefore be locally stiff yet easy to escape from, or locally soft yet protected by a large escape action.

$$\mu = V''(\psi_{\mathrm{eq}}) = a + 3b\psi_{\mathrm{eq}}^2 + 5c\psi_{\mathrm{eq}}^4$$
$$\tau_{\mathrm{rec}} = \frac{1}{\Gamma\mu} \tag{4}$$

For quasi-stationary small-noise dynamics, the escape-rate asymptotic depends on the applicable regime. $D_{\mathrm{eff}}$ denotes the effective fluctuation intensity in the approximate gradient model, whereas $\varepsilon$ is the formal small-noise scaling parameter in the general Freidlin-Wentzell action formulation:

$$k_{\mathrm{escape}} \propto \exp\left(-\frac{\Delta V}{D_{\mathrm{eff}}}\right) \quad \text{(gradient approximation)}$$
$$k_{\mathrm{escape}} \asymp \exp\left(-\frac{\Delta W}{\varepsilon}\right) \quad \text{(general small–noise action)} \tag{5}$$

Spontaneous transition rates identify only combinations such as $\Delta V/D_{\mathrm{eff}}$ in the gradient case or $\Delta W/\varepsilon$ in a general small-noise action formulation. These noise scales should not be equated unless a common normalization and diffusion model are specified. Small perturbations primarily estimate displacement and recovery, whereas larger perturbations probe empirical escape. The resulting dissociation is central: rapid local recovery can coexist with frequent full state escape, and slow local recovery can coexist with infrequent escape. Parameterized landscape claims must outperform model-free recovery slopes and escape hazards measured on the same graded perturbation scale.

## 3.5 Spatial recruitment

The Ginzburg term adds a spatial prediction absent from scalar sleep-onset models. In a local Gaussian approximation, the latent correlation length is

$$\xi \approx \sqrt{\frac{\kappa}{\mu}} \tag{6}$$

This relation concerns the latent ordering field, not raw scalp coherence. Reference choice, volume conduction, and the lead field can inflate scalp synchrony. Tests of $\xi$ therefore require source reconstruction, surface-Laplacian or current-source-density methods, MEG, high-density EEG, or intracranial recordings. At the N2-to-N3 boundary, the hypothesis is that local slow waves either recruit progressively broader cortical territories or fail to do so [52-55].

## 3.6 Candidate estimation, discriminant validity, and fallback coordinates

A candidate estimator for $\phi$ should be specified before transition testing. One defensible starting point is a multilevel confirmatory-factor or state-space model with prespecified indicators: slow-oscillation density or 0.5-1 Hz dominance, inverse signal complexity, and source-space spatial coordination. Indicators should be standardized within a defined reference condition, with age, cycle, montage, and pathology included as explicit moderators. A one-factor model should be compared with two- or three-factor alternatives and nonlinear embeddings using subject-held-out likelihood, loading stability, transport across recording conditions, and predictive performance.

Validation requires configural and at least partial metric/scalar invariance rather than unrealistic equality of every loading and intercept across age, montage, and pathology [64]. A small set of anchor indicators must retain stable meaning, while noninvariant indicators are modeled explicitly. Discriminant validity is essential: $\phi$ must provide a modest but reproducible gain in held-out prediction of a boundary trajectory, perturbation response, or outcome after conditioning on stage labels, N3 percentage, and the strongest individual component indicator. The expected gain may arise from denoising and invariant aggregation; superiority over all component indicators simultaneously is not required. If a scalar factor is inadequate, the fallback is a boundary-specific low-dimensional vector, for example separate coordinates for slow-oscillation organization, complexity, spatial recruitment, and event probability. The scalar-field implementation is withdrawn only when even partial invariance, transport, and discriminant prediction fail; the operation-by-location taxonomy can remain with vector or model-free coordinates.

## 3.7 Observation model, estimation, and circularity controls

The physiological state, the scored stage, and the subjective report are separate levels. A minimal generative observation model is

$$y_i(t) = \int_\Omega L_i(\mathbf{r})\Lambda_i\psi(\mathbf{r},t)\,d^d r + \varepsilon_i(t)$$
$$s_n = G(\{y(t)\}_n) \tag{7}$$
$$r_n = H(\{q(t)\}_{I_n},\ \text{memory},\ \text{expectation}) + \nu_n$$

Here $y_i$ is an observed channel or derived physiological feature, $L_i$ is the lead-field or sampling kernel, $\Lambda_i$ is a measurement loading, $s_n$ is the rule-based 30-s stage label, and $r_n$ is subjective report over interval $I_n$. In practice, $G$ is fixed by the declared scoring standard and scorer protocol. $L_i$ is fixed from the sensor geometry or source model when available; $\Lambda_i$ and the covariance of $\varepsilon_i$ are estimated in a hierarchical measurement model or calibrated on held-out segments. $H$ can be estimated with repeated diary reports, forced awakenings, or confidence-rated sleep/wake probes using an ordinal or probabilistic report model. Observation-layer dissociation is supported only when the report model adds reproducible prediction after objective transition instability, scoring uncertainty, local EEG, and memory-related covariates are modeled.

The latent coordinate must be prespecified and tested for partial measurement invariance and transport across subjects, ages, cycles, and montages. A feature used to define a boundary cannot then be used to prove a discontinuity at that boundary. Scored stages should identify candidate windows; transition time and class should be inferred from held-out continuous features or cross-fitting. Driven smooth trajectories, rounded bifurcation models, scoring-induced changes, hidden-state switches, stochastic escape, coexistence, and mixed models should be compared rather than assumed. In sleep-state misperception, objective instability and observation-layer error should be fitted jointly; the separate observation mapping is retained only when it improves cross-night prediction of report after objective dynamics and scoring uncertainty are included.

## 3.8 Multidimensional REM-to-wake dynamics

REM and wake can both show low cortical ordering while differing sharply in sensory coupling, motor access, eye-movement organization, autonomic patterning, and arousal regulation. We therefore use a normalized multidimensional state vector. Standard PSG directly informs $\phi$, $m$, and $e$; autonomic recordings provide proxies for part of $\gamma$; and $\rho$ generally requires evoked-response or awakening-probe data. Aminergic tone is not directly observable in routine human PSG and is therefore omitted from the minimal operational vector or represented only through declared proxies.

$$\dot{\mathbf{q}} = -M_s\nabla_q U(\mathbf{q};\lambda) + M_a\nabla_q U(\mathbf{q};\lambda)$$
$$+ f_{\text{cyc}}(\mathbf{q},z;\lambda) + \boldsymbol{\zeta}(t) \tag{8}$$
$$M_s = M_s^T \succ 0; \qquad M_a = -M_a^T$$

The minimal operational vector is $q_{\text{obs}} = [\phi, m, e, \gamma_{\text{proxy}}]^{\text{T}}$, with $\rho$ added when responsiveness probes are available. Local REM recovery is represented by a dissipative component $-M_s\nabla_q U$, where $M_s$ =

$M_s^T$ is positive definite. A reactive component $M_a \nabla_q U$, with $M_a = -M_a^T$, permits circulation along quasipotential contours, while a non-gradient term $f_{cyc}(q, z; \lambda)$ represents adaptation or coupling to an ultradian oscillator $z$. The full equation is not intended to fit a five-dimensional nonlinear SDE from one routine PSG night. Near a single REM epoch, $M_a$ and $f_{cyc}$ may be negligible and the local dissipative model estimates recovery. Whole-night NREM-REM sequencing requires the reactive or oscillatory terms and cannot be generated by a static symmetric-gradient model. In the present formulation, $z$ is imported rather than generated: the model describes how local REM-state dynamics couple to an ultradian rhythm, not an autonomous derivation of that rhythm itself.

## 3.9 Identifiability and baseline models

Most coefficients are identifiable only as effective combinations. Passive recovery identifies $\Gamma\mu$ rather than $\Gamma$ and $\mu$ separately; spontaneous gradient escape identifies $\Delta V/D_{eff}$, while general non-gradient escape identifies an effective action-to-noise scale such as $\Delta W/\varepsilon$; $h(t)$ can be confounded with asymmetric state-dependent noise; and pooled trajectories can mimic bistability. Standard PSG supports only a reduced REM vector and local recovery proxies. Multimodal measurements, controlled perturbations, matched forward-reverse trajectories, residual-current tests, strong priors, or repeated observations are required to reduce these confounds.

The Landau-Ginzburg family should be compared against simpler baselines: generalized additive or sigmoid trajectories, change-point models, transition matrices, hidden Markov or semi-Markov models, and boundary-specific neural-mass models. The more complex model is justified only if it improves held-out prediction or yields reproducible perturbation-response distinctions. The formalism is phenomenological; it does not derive coefficients from orexin, GABA, acetylcholine, thalamocortical loops, or cortical excitation-inhibition balance [43-45,61].

*Table 2. Notation, empirical status, and interpretive cautions at a glance.*

| Quantity | Role | Empirical status | Candidate estimator | Key caution |
|---|---|---|---|---|
| $\phi$, $\psi$, $\phi_0$ | Latent composite NREM ordering coordinate, centered at a reference | Latent; normalization prespecified | Latent-variable model using slow oscillations, complexity, and spatial organization | Indicators are correlated but not interchangeable |
| $a$, $b$, $c$ | Local polynomial coefficients and transition character | Fitted effective parameters | Generative model comparison across transition windows | Do not map one-to-one onto neurotransmitters |
| $\mu$ | Local curvature or stiffness | Derived fitted quantity | Small-perturbation displacement and recovery | Not equivalent to barrier height or stage duration |
| $\Delta V$ / $\Delta W$ | Gradient potential barrier / non-gradient escape action | Latent fitted quantity | Graded escape data, drift/noise model, or minimum-action estimation | $\Delta V$ is valid only for an approximate gradient model; passive hazard identifies an action/noise combination |
| Relative well depth | Preference between competing minima | Latent fitted quantity | Matched forward-reverse occupancy under comparable controls | Not the same as local curvature |

| Quantity | Role | Empirical status | Candidate estimator | Key caution |
|---|---|---|---|---|
| $h$ | Bias or tilt | Latent fitted quantity with physiological covariates | Circadian phase, arousal, sensory context, fitted asymmetry | Confounded with asymmetric noise and hidden drift |
| $D$, $D_{eff}$, $\varepsilon / \eta$ | Effective fluctuation intensity, covariance, and small-noise scaling | Latent residual process / asymptotic scaling | Noise color, residual covariance, dwell and escape statistics | $D_{eff}$ and $\varepsilon$ belong to different model normalizations and should not be equated; symmetric noise does not selectively favor wake |
| $\Gamma$ | Relaxation coefficient | Latent; usually identifiable only with $\mu$ | Perturbation recovery under known displacement | Passive data often identify $\Gamma\mu$ only |
| $\kappa$, $\xi$ | Spatial coupling and latent correlation scale | $\kappa$ latent; $\xi$ derived with spatial proxy | Source-space correlation and traveling-wave recruitment | Scalp coherence is not $\xi$ |
| **$q_{obs}$ and $q$** | Reduced observable and expanded REM/wake state vectors | Partly observed, partly proxy-defined | EEG, EOG, chin EMG, autonomic signals; responsiveness probes when available | Aminergic tone is not directly observed in routine human PSG; fit only an identifiable subset |
| $y$, $s$, $r$ | Continuous physiology, scored stage, subjective report | Observed / rule-generated | PSG/EEG, scoring, diary or probe | Disagreement can arise in the observation layer |

## 3.10 Operational granularity and minimum data requirements

The operational taxonomy is coarser than the full parameter vocabulary. Routine recordings can classify the failed operation and its location and can estimate transition timing, dwell distributions, reversal rates, and some effective combinations. Individual assignment to stiffness, barrier, bias, noise, relaxation, or coupling requires the protocol-specific evidence summarized below. Parameter-level labels should remain unassigned when the minimum design is absent.

*Table 3. Minimum data required for operational dynamical distinctions and fallback coordinates.*

| Target distinction | Minimum data or protocol | Recoverable inference | Operational status |
|---|---|---|---|
| Boundary approach vs crossing | Presleep wake plus continuous PSG; sleep history and circadian context | Distance/progression to boundary, transition time, and candidate crossing class | Potentially operational with routine PSG; causal attribution remains limited |
| Recovery rate $\Gamma\mu$ | Event-locked spontaneous or controlled small perturbations | Combined local recovery rate and displacement response | Estimable as an effective combination; $\mu$ and $\Gamma$ remain confounded |
| Escape propensity $\Delta V/D$ or $\Delta W/\varepsilon$ | Repeated dwells or multi-night transitions under approximate stationarity; explicit drift/noise model | Combined empirical escape propensity or action/noise scale | Effective combination only; gradient barrier requires evidence for approximate detailed balance |
| Curvature $\mu$ vs relaxation $\Gamma$ | Calibrated perturbation amplitude plus dynamical priors or multimodal time-constant measurements | Separate restoring geometry from kinetic relaxation | Forward-looking research target |

| Target distinction | Minimum data or protocol | Recoverable inference | Operational status |
|---|---|---|---|
| Gradient barrier $\Delta V$ vs noise $D$ | Graded perturbations, residual-noise model, and repeated escape trials | Separate gradient barrier contribution from fluctuation intensity | Forward-looking; use $\Delta W$ or hazard when non-gradient currents are present |
| Bias $h$ vs asymmetric noise | Matched forward-reverse trajectories with circadian, sensory, and autonomic covariates | Landscape tilt versus state-dependent fluctuation asymmetry | Research-grade; usually indeterminate in passive PSG |
| Spatial coupling $\kappa$ and latent $\xi$ | High-density EEG with source/CSD methods, MEG, or intracranial data | Local-to-global recruitment and latent correlation scale | Research-grade; raw scalp coherence is insufficient |
| Reduced REM vector and reactive dynamics | EEG, EOG, chin EMG, autonomic signals; responsiveness probes; multi-cycle data | Observable REM/wake trajectory, local recovery, and evidence for rotational/oscillatory drift | Research-grade; do not fit unobservable aminergic components from routine PSG |
| Scalar $\phi$ vs boundary-specific vector | Multi-night multimodal indicators; factor and predictive model comparison | Adequacy of one latent ordering dimension after stage/N3 adjustment | Use partial invariance and discriminant validity; fall back to a vector when scalar transport fails |
| Causal status | Longitudinal temporal ordering, alternative-cause assessment, mediation, or intervention | Primary, secondary, maintaining, or biomarker role | Undetermined by default in cross-sectional PSG |

Worked example. A single-night PSG may show prolonged wake-to-sleep latency and a reproducible transition trajectory, allowing classification at the coarse level of control-path progression or boundary crossing. Passive data may estimate a combined recovery rate or empirical escape propensity. A claim of specifically low curvature, low gradient barrier, or low non-gradient escape action requires calibrated perturbations, an explicit drift/noise model, and repeated responses. The coarse classification should also be withheld when a driven smooth trajectory and conventional covariates predict the event as well as the dynamical alternatives. This falsifying branch prevents the framework from forcing every complaint into a preferred operation.

## 4. A multiaxial dynamical taxonomy

A coherent research ontology should keep analytically different entities on separate dimensions. The proposed framework therefore uses three independent descriptive axes instead of collapsing candidate operations, stage locations, causal interpretations, and observation-layer discrepancies into one informal phenotype list. A patient profile is a composition of a dynamical operation, a stage or boundary, and a causal status. The matrix is a grammar for constructing hypotheses, not a claim that every cross-product cell exists, is independent, or defines a disease. Fine parameter hypotheses are nested within coarse operation-location descriptions and require the data standards in Table 3. Established clinical and physiological variables remain modifier/covariate dimensions.

## 4.1 Axis I: dynamical operation

The operation axis contains six classes: control-path progression; boundary crossing; post-transition stabilization; spatial recruitment; architectural sequencing; and observation mapping. These operation classes are the operational taxonomic level. Wake-state curvature, barrier height, bias, or noise are competing mechanistic explanations within boundary crossing rather than parallel diagnoses. Wake-attractor persistence and sleep-onset transition failure are therefore not separate classes. Likewise, unstable N1 and reduced spindle emergence must be distinguished by testing whether N2-event abnormalities persist after preceding-state stability is controlled.

## 4.2 Axis II: stage or boundary

The location axis specifies Wake → N1, N1 → N2, N2 → N3/SWS, NREM ↔ REM, sleep → wake, or the whole-night transition network. The location does not determine the mechanism. The same operation, such as low barrier escape, may occur in N1, N3, REM, or late-night N2; and the same location can express different operations, such as failure to enter N3 versus failure to sustain it.

## 4.3 Axis III: causal status

Every dynamical signature has a potential causal role: primary defect, secondary architectural consequence, maintaining mechanism, associated biomarker, or undetermined. Cross-sectional PSG usually supports only the undetermined category, together with exclusion of obvious event-linked secondary causes. Primary or maintaining assignments require longitudinal temporal precedence, alternative-cause assessment, and preferably mediation or intervention evidence showing that modification of the feature changes symptoms or vulnerability. The same REM fragmentation, reduced slow-wave organization, or high-frequency EEG feature may occupy different roles in different patients.

## 4.4 Modifier and covariate dimensions

Objective sleep duration is treated as a severity or exposure dimension that may cut across control-path, maintenance, terminal-arousal, and whole-night profiles rather than as an additional dynamical class [36-39]. Other modifier/covariate dimensions include chronicity, age, circadian phase, hyperarousal burden, comorbidity, medication exposure, and night-to-night variability. They are conceptually separate from the three classification axes but are not assumed to be statistically independent: they must be entered as covariates, moderators, or matching variables when dynamical quantities are estimated.

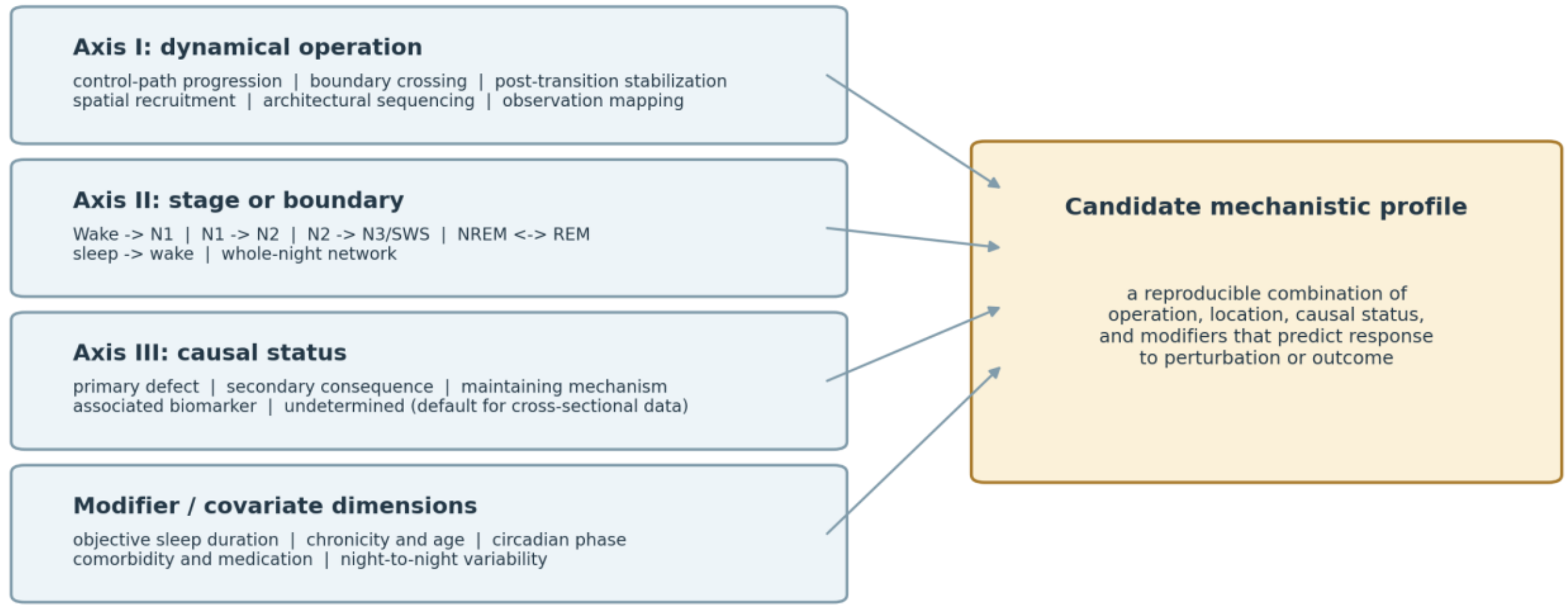

*Figure 1. Multiaxial architecture of the proposed framework. The arrows show analytical combination of dimensions into a candidate profile; they do not represent a causal chain. Causal direction is tested separately, and causal status remains undetermined by default in cross-sectional data.*

## 4.5 How the axes generate clinically recognizable profiles

At sleep onset, three profiles that can produce similar latency should be separated provisionally. Control-path underdrive means that homeostatic or circadian variables do not bring the system near the boundary. Boundary-crossing failure means that the boundary is approached but a driven smooth null, a rounded bifurcation model, or a stochastic switching model must still be distinguished. Failure of early-NREM consolidation begins after the first Wake → N1 crossing and is expressed by short N1 dwells, repeated W ↔ N1 reversals, or failure to accumulate N2 organization. These alternatives predict different presleep trajectories and perturbation responses, but no fold label is assigned without ruling out the control-parameter ramp and demonstrating model-specific evidence.

A primary N2 organization deficit should be assigned only when reduced spindle or K-complex probability, coupling, or spatial organization persists after preceding N1 stability and total N2 duration are controlled [32,50,51]. Otherwise, weak N2 graphoelements may be secondary to repeated interruption of early NREM. This hierarchical rule prevents state-stabilization failure and N2-event dysfunction from being counted as independent parallel disorders when one causes the other.

Sleep-maintenance complaints can likewise reflect at least two dynamical patterns. A low gradient barrier or low non-gradient escape action predicts frequent state exits that need not be preceded by slow recovery. A soft occupied state predicts a large response and prolonged recovery after a small perturbation. Respiratory events, movements, pain, vasomotor symptoms, environmental stimuli, and medication effects can create either pattern as a secondary abnormality, so intrinsic insomnia should be inferred only after these inputs are modeled.

Reduced deep sleep can arise from failure of N2-to-N3 spatial recruitment or from erosion after N3 has been reached. Recruitment failure predicts local slow-wave events without normal growth of source-space correlation and broad propagation. Erosion predicts relatively normal entry followed by short high-ordering dwells, exaggerated perturbation sensitivity, or early return to N2 or wake. Conventional N3 percentage cannot distinguish these possibilities [21,31,32].

REM complaints require a reduced observable state vector, expanded with responsiveness probes when available. Abnormal entry, phasic-tonic organization, maintenance, and REM-to-wake escape should be separated after controlling source stage, cycle number, prior NREM, clock time, depression, medication, and sleep-disordered breathing [33,34]. Local REM recovery may be approximated dissipatively, whereas whole-night sequencing requires reactive or oscillatory dynamics. Premature terminal arousal is a late-night combination of control-path and boundary-crossing change, while sleep-state misperception is assigned to the observation axis unless objective instability is demonstrated.

**Examples generated by crossing dynamical operation with stage or boundary**

| | Wake -> N1 | N1 -> N2 | N2 -> N3/SWS | NREM <-> REM | Sleep -> wake | Whole night |
|---|---|---|---|---|---|---|
| **Control path** | homeostatic underdrive / circadian bias | | | | | |
| **Boundary crossing** | wake-state persistence | | | abnormal REM entry or exit | premature terminal arousal | |
| **State stabilization** | early-NREM instability | failure of N2 consolidation | SWS erosion | REM-state instability | | |
| **Spatial recruitment** | | spindle / K-complex organization | SWS recruitment deficit | | | |
| **Architectural sequencing** | | | | | | cycle / transition-network dysregulation |
| **Observation mapping** | sleep-onset misperception | local sleep / report mismatch | | REM experience / physiology mismatch | | multi-night report mapping |

*Filled cells show illustrative profiles, not an exhaustive list. Blank cells are not presumed impossible. Causal status is assigned on a separate axis.*

*Figure 2. Compositional operation-by-location matrix. Filled cells are worked examples generated by crossing an operation with a stage or boundary; they are not an exhaustive list, prevalence claim, or assertion that all cells are independent biological entities. Blank cells are neither proposed nor ruled out. Causal status and modifier/covariate dimensions are assigned separately.*

## 4.6 Candidate profiles generated by the axes

The profiles below are selected shorthand mechanistic phenotypes generated by the compositional grammar. They are not validated endotypes, mutually exclusive disorders, or an enumeration of the full Cartesian product. A cell is retained only when it yields a distinctive measurement or falsification strategy. "N1 trapping" is avoided because the proposed early-NREM phenotype often consists of repeated escape from N1 rather than prolonged residence; the preferred term is failure of early-NREM consolidation or N1 instability.

*Table 4. Candidate mechanistic profiles expressed in the multiaxial framework.*

| Candidate profile | Operation | Location | Causal status | Evidence | Major confounds | Decisive / falsifying test |
|---|---|---|---|---|---|---|
| Homeostatic underdrive / circadian wake bias | Control-path progression | Wake → N1 | Primary, precipitating, or secondary | Motivating physiology | Schedule, naps, phase disorder, behavior | Normal approach/slowing despite failed sleep rejects a pure control-path account |
| Wake-to-sleep crossing failure | Boundary crossing: driven rounding, recovery, escape, or bistability | Wake → N1 | Primary or maintaining | Strong sleep-onset evidence; mechanism not identified; insomnia enrichment unproven | Hidden drift, scoring delay, conditioning | Driven sigmoid vs rounded/bistable/escape comparison; no predictive gain rejects the mechanistic label |
| Failure of early-NREM consolidation | Post-transition stabilization / easy escape | Wake/N1 and N1 → N2 | Primary, maintaining, or secondary | Hypothesis-motivating transition/perturbation findings | Sensory events, apnea, movements, scoring uncertainty | W↔N1 reversals persist after exogenous arousals and scoring uncertainty are modeled |
| Primary N2 event-organization deficit | Spatial/event organization | N1 → N2 | Primary, secondary, or biomarker | Compatible NREM microstructure | Unstable N1, detector threshold, medication | Deficit persists after N1 stability and N2 duration are controlled |
| Sleep-maintenance escape / recovery | Stabilization and boundary crossing | N2, N3, REM → wake | Primary, secondary, or maintaining | Compatible perturbation/transition evidence | Apnea, movements, pain, environment | Recovery and escape dissociate under one graded perturbation scale and outperform model-free measures |
| SWS recruitment failure | Spatial recruitment | N2 → N3/SWS | Primary, secondary, or biomarker | Hypothesis-motivating slow-wave findings | Age, homeostatic pressure, scoring threshold, lead field | Local slow waves fail to recruit broader source-space territories |
| SWS erosion | Post-transition stabilization | Within N3/SWS or N3 → N2/wake | Primary, secondary, or maintaining | Compatible N3 perturbation evidence | Failure to enter N3, respiratory/autonomic arousal | Normal N3 entry with short dwell or exaggerated escape; absent dissociation rejects subtype |
| REM multidimensional instability | Boundary crossing / stabilization | NREM ↔ REM and REM → wake | Primary, secondary, maintaining, or biomarker | Compatible REM microstructure | Cycle, depression, medication, apnea, source-stage pooling | Abnormality persists after macroarchitecture, source stage, prior NREM, and clock time are controlled |
| Premature terminal arousal | Control path plus boundary crossing | Late-night N2/REM → wake | Primary or secondary | Motivating circadian/clinical pattern | Phase advance, depression, age, environment | Late-night hazard remains abnormal after phase, Process *S*, and comorbidity adjustment |
| Observation-layer dissociation | Observation mapping | Any stage; often light NREM | Primary/secondary mapping defect or biomarker | Compatible multi-night/evoked evidence | Objective instability, recall bias, scoring rules | Discrepancy remains after objective transition instability is modeled |
| Whole-night architectural dysregulation | Architectural sequencing | Whole-night network | Primary, secondary, maintaining, or summary biomarker | Motivating transition-network evidence | Single-night variability, pooled stages, exogenous events | Cross-night pattern is reproducible and predicts beyond TST and WASO |

# 5. Evidence map

The evidence should be interpreted through an explicit hierarchy. A mechanism-discriminating signature directly estimates a transition property and rules out major driven, scoring, and simpler-model alternatives. A finding strongly consistent with a proposed mechanism fits the predicted pattern but does not uniquely identify it. A hypothesis-motivating finding establishes a relevant abnormality without locating it in the proposed geometry. A speculative mapping extrapolates beyond the design of the cited study. No insomnia domain currently satisfies the strict mechanism-discriminating standard; sleep onset provides the strongest boundary-level evidence, while the other operations remain hypothesis-generating extensions.

## 5.1 Strongest boundary-level evidence: sleep onset

Sleep onset provides the strongest current boundary-level evidence. Li and colleagues identified a low-dimensional coordinate with a predictable bifurcation-like trajectory and passive early-warning signals in two large cohorts [40]. Hu and colleagues fitted a stochastic bistable model that reproduced wake-sleep flickering and related drift/noise parameters to sleepiness [41]. These results are strongly consistent with bifurcation-like or bistable dynamics. They do not yet distinguish intrinsic critical slowing from slowing induced by the deterministic Process-*S*/Process-*C* ramp, establish small-field bistability, demonstrate hysteresis, or show enrichment of a particular mechanism in insomnia. Fold language is therefore reserved for analyses that outperform a driven smooth null and identify branch loss or matched-control bistability.

## 5.2 Findings consistent with selected components

A multicenter CBT-I study found dissociable changes in the NREM delta/beta ratio and transition-matrix stability [30]. This is consistent with the framework's separation of cortical activation from macrostate stability, but it does not identify curvature, barrier height, bias, or noise. An auditory perturbation study found greater evoked complexity and P300 responses, especially in N3, with altered intermediate-window coupling [31]. These findings are compatible with reduced perturbation resistance or containment, but graded perturbations are required to distinguish a soft state from a low barrier.

Reduced 0.5-1 Hz slow-oscillation power in older adults with insomnia and reduced spindle and slow-oscillation density in a broader adult sample motivate N2/SWS organization domains [21,32]. They do not distinguish reduced local ordering, weak spatial recruitment, or low state stability, and null findings in other samples argue against a universal deficit [19].

Large-sample REM findings - increased REM arousals and awakenings, REM density, and phasic REM proportion, including in participants without conventional macrostructural abnormalities - motivate treatment of REM microstructure as a distinct domain [33]. They are compatible with REM instability but do not uniquely identify reduced REM curvature, lowered barrier height, altered bias, or multivariate noise. Restless REM has also been linked to impaired overnight amygdala adaptation [34].

Sleep-state misperception studies show altered NREM spectral activity, evoked processing, reproducible multi-night discrepancy patterns, and associations with N3 proportion [23-28]. These findings motivate an observation-layer model but do not establish whether the dominant source is local wake-like physiology, sensory access, memory encoding, expectations, or objective transition instability.

### 5.3 Null, heterogeneous, and disconfirming evidence

Several findings constrain broad claims. Some carefully characterized insomnia samples show little or no group-level waking or NREM spectral difference [19], and meta-analytic PSG effects are heterogeneous rather than diagnostic at the individual level [29]. Slow-oscillation, spindle, and conventional N3 abnormalities are absent in subsets, so N2/SWS profiles cannot be universal [19,21,29,32]. Misperception studies identify multiple discrepancy patterns rather than one observation-layer physiology [25-28]. REM microstructural differences can coexist with near-normal macroarchitecture [33], which motivates finer measurement but also shows that stage totals and microstructure may dissociate. No study has yet demonstrated stable operation-location profiles across multiple nights or shown that the proposed parameters outperform simpler model-free measures. These nulls and gaps constrain prevalence and mechanism claims rather than serving as confirmatory evidence for the framework.

### 5.4 Modifier/covariate severity and exposure dimensions

Insomnia with objective short sleep duration is associated with cardiometabolic and neuropsychological risk and may represent a biologically severe phenotype [36-39]. Within the present framework it is a modifier/covariate dimension: short sleep may result from persistent wake bias, repeated maintenance escape, premature terminal arousal, or whole-night dysregulation. The decisive question is whether objective duration modifies risk independently of the inferred profile and whether the profile explains heterogeneity within short-sleep and normal-duration groups.

**Box 1. Speculative future hypothesis: a within-N3 mixed regime**A highly consolidated within-N3 state may contain a distinguishable high-coherence subregime whose entry or maintenance fails in some patients. A mixed continuous-plus-switch or tricritical-like normal form is one mathematical idealization [42]. This proposal is speculative and nonessential to the taxonomy. It should be abandoned if independently constructed latent features do not identify a reproducible substate with derivative change, bimodality, matched path dependence, or localized recruitment followed by propagation.

## 6. Predictions, taxonomy utility, and model falsification

### 6.1 Transition-centered analysis

The unit of analysis should be a transition event or state dwell rather than a 30-s epoch treated as ground truth. Expert labels may identify candidate windows, but transition time should be estimated independently from continuous features. At sleep onset, competing models must include a driven smooth sigmoid whose control variable is a Process-*S*/Process-*C* ramp, a rounded bifurcation model with estimated bias/rounding, stochastic bistable or escape models, change-point or hidden-state switches, and simpler nonparametric trajectories. Subjects, not epochs, should be held out in cross-validation, and null boundaries should be jittered within matched sleep-cycle segments.

Passive variance, autocorrelation, and detrended fluctuation measures can all increase under a deterministic control-parameter ramp and therefore cannot identify a fold by themselves. The preferred test is controlled perturbation-recovery at matched homeostatic pressure, circadian phase, clock time, prior-state duration, and sensory conditions. When only passive data are available, the driven-ramp null must be fitted explicitly, and claims should be limited to bifurcation-like or bistable consistency unless the rounded dynamical model improves subject-held-out prediction and shows model-specific structure.

## 6.2 Perturbation-response dissociation

Small perturbations should estimate local displacement and recovery, whereas larger perturbations on the same calibrated scale should estimate escape hazard. The framework predicts rapid recovery with frequent escape in some profiles and slow recovery with infrequent escape in others. This is a useful dynamical dissociation only if fitted recovery/escape parameters predict held-out physiology or clinical outcomes better than the model-free recovery slope and escape hazard themselves; otherwise the Landau parameterization adds terminology rather than information.

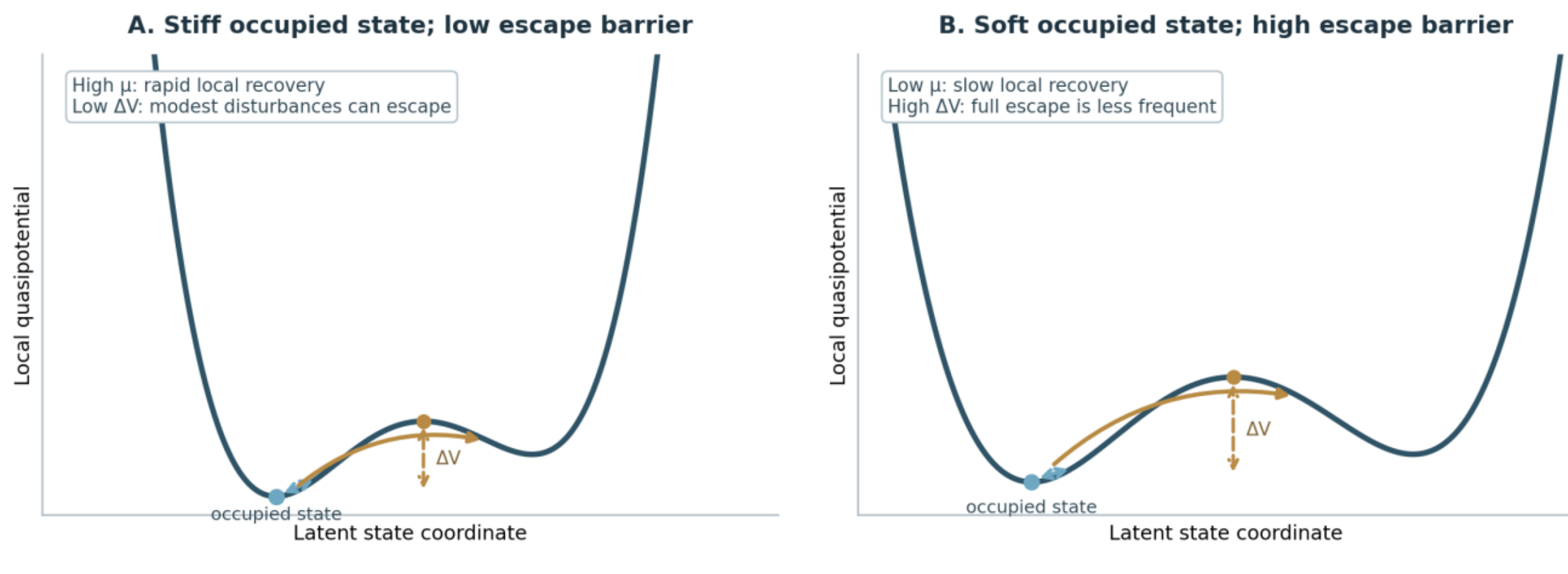

*Figure 3. Gradient-idealized dissociation between local stiffness and escape. With at least two free shape parameters, $\mu$ and $\Delta V$ can vary independently: a stiff/low-barrier profile recovers quickly but escapes readily, whereas a soft/high-barrier profile recovers slowly but rarely escapes. In non-gradient dynamics, $\Delta W$ or empirical hazard replaces $\Delta V$. Fitted parameters must outperform model-free recovery and escape measures. Schematic only.*

## 6.3 Cross-night reproducibility and predictive superiority

If these profiles are true endotypes, fitted parameter combinations and perturbation-response signatures should be more reproducible within individuals than conventional night-to-night stage percentages. They should also predict treatment response, relapse, perturbation vulnerability, or subjective-objective discrepancy beyond sleep-onset latency, WASO, total sleep time, N3 percentage, and symptom severity. Failure to add prediction would reduce the framework to descriptive terminology.

## 6.4 Taxonomy utility criteria

The three-axis taxonomy is an ontology rather than a physical law. It earns its place if trained raters can apply operation-location-causal labels reliably, if the labels improve communication or trial stratification, or if they add held-out prediction beyond conventional onset/maintenance/terminal-awakening/nonrestorative subtyping and standard covariates. If it produces poor inter-rater agreement, excessive unresolvable ambiguity, or no incremental predictive utility, the ontology should be simplified or retired even when an individual dynamical model remains mathematically adequate.

## 6.5 Model and profile falsification

The scalar-field implementation is rejected if $\phi$ lacks partial invariance, discriminant validity, and transport, or if vector and simpler models consistently predict better. A particular mechanistic profile is rejected when its required transition or perturbation signature is absent after confounds are modeled. The Landau-Ginzburg implementation is rejected for a boundary when a driven sigmoid, HMM/semi-Markov model, change-point model, or model-free perturbation description consistently wins on subject-held-out prediction. Failure of recovery and escape estimates to predict graded perturbation responses further rejects the proposed geometric interpretation. In those cases, landscape language may remain a communication metaphor but should not be treated as evidence for a physiological mechanism.

## 6.6 Near-term crux experiment

The most decisive currently feasible test is a preregistered reanalysis of an existing large sleep-onset cohort. The same continuous onset trajectories should be fitted with (i) a driven smooth sigmoid in which Process $S$ and circadian phase supply the ramp, (ii) a rounded bifurcation model with an estimated bias/rounding parameter, (iii) stochastic bistable or escape models, and (iv) hidden-state/change-point baselines. Model selection should use subject-held-out likelihood and calibration, with the transition-localizing features separated from the testing features. If the driven sigmoid predicts as well as or better than the rounded or bistable models, the manuscript's sleep-onset claims must be downgraded to descriptive state change and the top evidence tier remains empty. A reproducible predictive advantage for the dynamical alternatives would justify targeted perturbation studies and sharpen the framework's strongest empirical foothold.

*Table 5. Priority studies for testing model validity and taxonomy utility.*

| Priority analysis | Population / data | Primary outcome | Prediction | Falsifying result |
|---|---|---|---|---|
| Near-term crux: driven onset model comparison | Existing large cohorts with continuous presleep wake and sleep-history/circadian covariates | Subject-held-out driven sigmoid vs rounded bifurcation vs bistable/escape vs hidden-state models | A dynamical model adds reproducible predictive gain after the Process-*S*/Process-*C* ramp and scoring features are controlled | The driven sigmoid matches or outperforms; fold/bistability claims are withdrawn and onset remains descriptive |
| Early-NREM consolidation | Rapid drowsiness, high N1, repeated perceived non-sleep | W↔N1 dwell, N2-entry probability, perturbation recovery | A subset shows short N1 dwells, high reversal probability, or soft-state responses | N1 stability is normal after exogenous arousals and scoring uncertainty are controlled |
| N2 event organization | High-quality central/frontal EEG across multiple nights | Spindle/K-complex probability and coupling | A subset shows primary N2 organization deficit independent of N2 minutes and preceding N1 stability | Differences disappear after detector threshold, N1 stability, and medication are controlled |
| SWS spatial recruitment | High-density EEG/MEG or source-localized EEG | Source-space $\xi$, local onset, propagation | SWS-recruitment phenotype shows local slow waves with curtailed local-to-global recruitment | Spatial recruitment is normal despite reduced N3 or nonrestorative sleep |
| N3 perturbation resilience | Stage-locked graded auditory protocol | Resistance, containment, recovery, and escape | Recovery and escape dissociate and fitted parameters predict beyond model-free slope and hazard | One undifferentiated arousal response explains both, or model-free measures predict equally well |
| REM multidimensional dynamics | EEG/EOG/EMG/autonomic PSG | REM dwell, arousal hazard, $q$-vector trajectories, phasic-tonic structure | REM abnormality persists after macroarchitecture, cycle, source stage, and prior NREM controls | Differences disappear after those controls |
| Misperception observation model | Seven-plus nights with diaries and EEG/actigraphy | Latent state-to-report mapping | Discrepancy profile is separable from objective transition instability | Objective instability fully explains discrepancy and the separate observation model adds no prediction |
| Treatment-mechanism mapping | Pre/post CBT-I or pharmacotherapy | Change in control, recovery, escape, event organization, and symptoms | Treatments with similar total-sleep-time gains alter different dynamical domains | Dynamical variables are neither reproducible nor related to symptoms or treatment response |
| Scalar-coordinate validation | Multi-night PSG/hdEEG across ages, cycles, montages, and insomnia phenotypes | Factor structure, partial invariance, discriminant validity, and held-out prediction of $\phi$ | A stable scalar coordinate predicts beyond stage, N3%, and component indicators across conditions | Even partial invariance or discriminant validity fails; a vector model is required |
| Simulation-based identifiability | Synthetic data matched to PSG sampling, noise, and missingness | Recovery of $\Gamma\mu$, $\Delta V/D$, $h$/noise, and $\kappa/\xi$ under candidate protocols | The proposed minimum protocols recover their stated effective quantities without systematic bias | Nominal quantities remain unrecoverable or simpler models produce indistinguishable fits |
| Taxonomy utility | Clinician/researcher raters plus multi-night cohorts or trial datasets | Inter-rater reliability and incremental outcome prediction of operation-location-causal labels | The ontology improves communication, stratification, or prediction beyond conventional subtypes | Poor reliability, excessive undetermined labels, or no incremental utility warrants simplification or retirement |

# 7. Clinical interpretation and intervention hypotheses

The immediate value of the framework is clinical interpretability and research stratification, not actionability. It suggests augmenting a standard assessment by asking whether the dominant complaint is failure to become sleepy, failure to cross sleep onset, repeated early-stage reversals, nocturnal escape from established sleep, failure of N2 or SWS organization, REM fragmentation, premature terminal arousal, whole-night dysregulation, or a discrepancy between experienced and measured sleep. These patterns can guide differential measurement, but prospective evidence is required before they guide treatment selection.

CBT-I can alter several dimensions simultaneously: homeostatic pressure, conditioned wake bias, circadian regularity, and arousal. The dissociation between spectral and transition-matrix changes after CBT-I motivates multidomain measurement rather than a single mediator [30]. Pharmacologic agents that produce similar total sleep time may likewise differ in their effects on boundary crossing, local recovery, escape, N2 event probability, or REM organization. These are hypotheses for mechanism studies, not current prescribing rules.

Closed-loop neuromodulation is included solely as an illustrative example of hypothesis generation. In principle, stimulation might be timed to modify a wake-biased control trajectory, facilitate boundary crossing, stabilize newly entered NREM, promote spatial recruitment, or reduce perturbation vulnerability. Any such approach must preserve natural ultradian ordering and requires independent efficacy and safety testing, state-specific stopping rules, and careful control of sensory arousal and photosensitivity. The current framework supplies no validated target, product rationale, or dosing rule.

## 8. Limitations and boundary conditions

The framework is phenomenological. Its coefficients summarize low-dimensional state-space geometry and do not identify a unique circuit, neurotransmitter, or molecular mechanism. Different biological causes may produce similar effective geometry, and the same cause may alter several parameters. The model narrows hypotheses but does not replace mechanistic neuroscience.

The relaxational quasipotential is local and conditional on approximate gradient dynamics. A global scalar potential need not exist for the full sleep cycle because non-gradient probability currents, adaptation, delays, REM-on/REM-off oscillations, and Hopf-like components contribute to whole-night dynamics. In those regimes, escape must be described through a minimum-action quasipotential or model-free hazard, and sequencing requires reactive or oscillator terms. Process $S$, Process $C$, and neuromodulatory variables remain external controls to the local models rather than autonomously generated state variables.

A scalar NREM-ordering coordinate is a deliberate reduction and may fail when spindles, K-complexes, local slow waves, and high-frequency activation vary independently. REM and wake require a multidimensional vector. Measurement invariance, lead-field effects, finite-size rounding, colored noise, first-night effects, and stage-scoring rules can all manufacture or obscure apparent transition signatures.

The evidence base is sharply asymmetric. Sleep onset has the strongest boundary-level evidence, but even there a driven homeostatic-circadian ramp remains a serious alternative to intrinsic critical slowing. N2, SWS, REM, maintenance, sequencing, and perception assignments are compatible or hypothesis-motivating extensions, and explicit null findings constrain any universal claim. Group differences do not establish individual endotypes. Routine PSG generally supports coarse operation-location descriptions and effective combinations rather than fine parameter labels, while causal status is usually undetermined without longitudinal or interventional evidence.

Phase-transition-like language is retained only when operational criteria are specified. Passive slowing does not prove a fold when a driven ramp can generate the same pattern; bimodality does not prove coexistence; nonoverlapping paths should be called path dependence unless controls are matched; local slow waves do not prove nucleation; and spontaneous escape rates do not separately identify a gradient barrier, non-gradient action, and noise. Subjective sleep remains a core clinical outcome, and a physiological profile that does not explain distress, impairment, or outcome has limited value.

# Conclusion

Insomnia may be usefully conceptualized as a family of potentially separable failures in slow control trajectories, state-boundary crossing, post-transition stabilization, spatial recruitment, architectural sequencing, and the mapping between physiological and experienced sleep. The paper's principal contribution is a compositional three-axis ontology that keeps operation, location, and causal status separate. It is judged by pragmatic utility rather than physical truth. The Landau-Ginzburg formalism is an adopted, boundary-local scaffold whose gradient and non-gradient implementations are separately testable. Clinical and physiological modifiers are covariates rather than statistically independent axes.

The dynamical formulation is valuable only insofar as it generates distinctions that survive comparison with simpler alternatives: boundary approach versus crossing, local recovery versus escape, local slow-wave generation versus spatial recruitment, and latent state versus report. Published sleep-onset findings are strongly consistent with bifurcation-like or bistable dynamics but remain vulnerable to a driven smooth explanation; the other domains are hypothesis-generating. The immediate empirical priority is the preregistered driven-sigmoid versus rounded-bifurcation/bistable crux analysis, followed by latent-coordinate validation, simulation-based identifiability studies, and graded perturbations.

A useful dynamical taxonomy would complement diagnosis by asking: at which operation and boundary does the sleep process appear to fail, what causal role remains plausible, and does that profile improve communication, stratification, or prediction beyond conventional measures? A successful dynamical model must answer the additional question of whether its inferred geometry predicts perturbation response and outcome better than driven smooth, hidden-state, and model-free alternatives.

## Declarations

### Funding

No funding was received specifically for preparation of this theoretical manuscript.

### Conflicts of interest

Alexander Poltorak is affiliated with NeuroLight, Inc., a company developing technologies related to sleep monitoring and neuromodulation. The company may have intellectual-property and commercial interests related to sleep modulation. This manuscript is theoretical and does not establish a diagnostic biomarker or validated treatment protocol.

### Author contributions

Alexander Poltorak: conceptualization, theoretical framework, literature synthesis, writing, and revision.

### Data and code availability

No new clinical data were generated. No new empirical analysis was performed for this conceptual manuscript. The proposed validation analyses can be implemented in public or controlled-access polysomnography datasets, subject to their data-use requirements.

### Acknowledgments

The author thanks colleagues for discussions on sleep-stage dynamics, EEG transition analysis, and closed-loop neuromodulation.